\documentclass[preprints,article,accept,moreauthors,pdftex]{Definitions/mdpi}

\firstpage{1} 
\pubvolume{xx}
\issuenum{1}
\articlenumber{5}
\pubyear{2019}
\copyrightyear{2019}
\history{Published: 13/07/2020}

\usepackage{cleveref}
\usepackage{rotating}

\Title{Migration and Refugee Crisis: a Critical Analysis of Online Public Perception}

\Author{Isa Inuwa-Dutse$^{\dagger}$\orcidA{}, Mark Liptrott\orcidB{} and Ioannis Korkontzelos\orcidC{}}

\AuthorNames{Isa Inuwa-Dutse, Mark Liptrott and Ioannis Korkontzelos}

\address[1]{Department of Computer Science, Edge Hill University, Ormskirk, United Kingdom\\
            Email: I.Inuwa-Dutse@herts.ac.uk, Mark.Liptrott@edgehill.ac.uk, Yannis.Korkontzelos@edgehill.ac.uk}

\firstnote{Current affiliation: School of Engineering and Computer Science, University of Hertfordshire} 

\abstract{
The migration rate and the level of resentments towards migrants are an important issue in modern civilisation.
The infamous \textit{EU refugee crisis} caught many countries unprepared, leading to sporadic and rudimentary containment measures that, in turn, led to significant public discourse. 
Decades of offline data collected via traditional survey methods have been utilised earlier to understand public opinion to foster peaceful coexistence. 
Capturing and understanding online public opinion via social media is crucial towards a joint strategic regulation spanning safety, rights of migrants and cordial integration for economic prosperity. 
We present a analysis of opinions on migrants and refugees expressed by the users of a very popular social platform, Twitter. 
We analyse sentiment and the associated \textit{context of expressions} in a vast collection of tweets related to the \textit{EU refugee crisis}. 
Our study reveals a marginally higher proportion of negative sentiments vis-\'{a}-vis migrants and a large proportion of the negative sentiments is more reflected among the ordinary users. 
Users with many followers and \textit{non-governmental organisations (NGO)} tend to \textit{tweet} favourably about the topic, offsetting the distribution of negative sentiment. 
We opine that they can be encouraged to be more proactive in neutralising negative attitudes that may arise concerning similar incidences.}


\begin{document}

\section{Introduction}
\label{sec1:introduction}

Modern social media platforms such as Twitter\footnote{\url{twitter.com}} and Facebook\footnote{\url{facebook.com}} are very popular with the public for many reasons. 
Users of the platforms can generate and consume content simultaneously on a daily basis leading to various forms of information, e.g.~fads, opinions and breaking news, on various social phenomena. 
Users share information about virtually all aspects of their social life making social media platforms ideal for studying various aspects of social phenomena\footnote{Areas such as event detection, disaster monitoring and management, politics, healthcare, sports, have been extensively studied using social media data.}. 
These capabilities lead to vast amounts of resourceful data from diverse users \cite{chakraborty2016fashioning}. 
The platforms offer crucial avenues where users express views on global issues and can affect, and be useful in understanding how modern society functions \cite{clemencecarl}. 
In this study, we analyse how data from social media can be utilised to inform solutions to problems on migration and refugee crisis.

While migration has been around for a very long time and will continue, the present unprecedented migration rate is seen as one of the pressing issues challenging modern civilisation. 
For instance, the infamous \textit{EU refugee crisis} caught many countries unprepared. 
Most challenges common to mass migration refer to obligations on the part of host nations, e.g.~identifying genuine asylum seekers, managing with large numbers of refugees and the threats to security.
Challenges of this kind trigger all forms of sporadic and rudimentary measures to regulate or stop the inflow of migrants \cite{castles2000international}. 
For instance, at the height of the crisis, while some countries allowed migrants to enter, other countries either blocked entries, prepared remote holding camps or prevented migrants from forming local camps.
Migration and the cultural diversity of ethnic migrants have become major public issues causing much controversy among politicians and the general public for a long time \cite{coenders2003majorities}. 
This cultivates and attracts different opinions and resentments among the public, which cannot be ignored. 
Political undertones and hazy media reports \cite{fuchs2017social} often lead to antagonistic attitudes from the public \cite{gross2011migrations}. 
This often attracts the attention of the media reporting all forms of news with consequences for both migrants and the receiving communities. 
The traditional media have been considered to contribute to resentments and unwelcoming attitudes in the receiving communities through the creation of what was  termed \textit{create crisis mentality} \cite{esses2013uncertainty}.

Noting the impact of \textit{changing environment} in understanding problems and proffering effective solutions (see Figure \ref{fig1:migration-policy}), this study recognises that effective solutions to problems affecting the well-being of society cannot afford to ignore input/perspectives from the rich data available in social platforms. 
Inclusion of aspects within social media related to public perception will ensure a holistic approach by recognising the role of the online communities in confronting the challenges. 
An effective strategy that will improve and accelerate broader integration and coexistence with the hosting community is the desired outcome resulting in economic growth and social harmony. 

In order to produce this understanding and to provide a holistic perspective derived from online communities, this study will:
\begin{enumerate}[leftmargin=*,labelsep=4.9mm]
    \item quantify the proportions of sentiment/perception of online users on the subject matter. 
    \item measure the sentiment of the public across global regions.
    \item analyse the discussion topics and their respective sentiment.
    \item investigate the role of influential users,i.e.~users with many followers on Twitter, such as politicians or media publishers.
	\item identify relevant aspects to consider for a more inclusive regulatory framework with respect to migration and refugees.    
\end{enumerate}

Joint and strategic efforts from various stakeholders will ensure effective regulation \cite{sassen2005regulating}, safety and rights of migrants and peaceful coexistence with the receiving communities \cite{castles2000international}. 
Public opinion has been shown to influence and shape immigration policy \cite{lahav2004public}, which in turn requires multi-perspective efforts from various stakeholders \cite{sassen2005regulating,unit2016measuring}. 
The goal of this study is to understand the polarisation of opinions among the public in online communities, specifically among the users of Twitter. 
The \textbf{contributions} of this study are the following:

\begin{enumerate}[leftmargin=*,labelsep=4.9mm]
    \item We provide a comprehensive study that critically analyses online public opinion about migration and the refugee crisis. 
    We examine the proportions of sentiments across different user groups - \textit{verified and unverified} and across \textit{continental regions}. 
    With the exception of the \textit{verified users group}, \textit{negative sentiments} dominate the online public opinion.
    \item We show how online users with social influence can help in improving peaceful engagement and balance the disproportionate ratio of \textit{negative} vs.~\textit{positive} comments.
    \item We analyse the discussion topics in the dataset according to the sentiment class and describe the nature of each topic.
    \item Insight from this research can be used to inform policies and legislative frameworks for the settlement and integration of migrants and refugees into the receiving communities. 
    The outcome will be of relevance to the \textit{European Monitoring Centre on Racism and Xenophobia (EurWORK)} to strengthen the integration of migrants to ensure peaceful coexistence between those groups and the local communities. 
    To that end, we propose vital areas to consider to neutralise negative impacts.
    \item Finally, the dataset will be made available to interested researchers in accordance with the Twitter sharing policy.
\end{enumerate}

The paper is structured as follows:  
Section \ref{sec2:background} provides some background information related to the subject matter. 
Section \ref{sec3:review} presents the related work and Section \ref{sec4:experiments} describes the method, the materials used and the experimental results. 
Section \ref{sec5:discussion} offers a detailed discussion of our findings and Section \ref{sec6:conclusion} concludes the study.

\section{Background}
\label{sec2:background}

This section outlines the framework upon which the study is based. 
We present a general discussion on social media, contemporary migration and frameworks for regulating migration.

\subsection{Social network} 

Today's social media platforms enable users to both generate and consume online contents leading to a complex information ecosystem. 
Users share information about virtually all aspects of their social life on the platforms, hence serving as a useful source of latest information, ideal for studying various aspects of social phenomena. 
These platforms enable users to generate and consume content simultaneously on a daily basis leading to various forms of information, making social media a rich source of longitudinal data from diverse users. 
Because users decide what to share and engage openly or privately, a huge amount of resourceful and self-policing data is available \cite{chakraborty2016fashioning}. 
Concerning migrants and refugees, the recognition of a \textit{changing environment} and of citizen's opinions play a pivotal role in ensuring effective regulation. 
Figure \ref{fig1:migration-policy} summarised key areas for consideration in this respect. 
In this study, we recognise that effective solutions to problems affecting the well-being of the present society cannot afford to ignore how the online public view the problem. 
Social media platforms affect and offer an understanding of how modern society functions \cite{clemencecarl}. 
This increasing amount of social media data allows to understand public perception on migrants and refugees, and this perception complements insights from studies based on the \textit{Eurobarometer data}.

\subsection{Contemporary migration}

Current migration contributes to a significant percentage of population growth. 
The number of migrants globally amounts to 258 million in 2017 and is increasing at a rapid rate of 49\% since 2000 growing faster than global population \cite{desa2017united}. 
Many reasons related to push and pull factors\footnote{Push factors relate to conflicts or life-threatening events, whereas pull factors are attributed to the urge for greener pastures \cite{gross2011migrations}.} have been cited as responsible for exacerbating the problem. 
A substantial proportion of contemporary migrants are classified as refugees, mainly due to war, persecution, injustice, exclusion, environmental pressure, competition for scarce resources and economic hardship \cite{trends2009refugees}. 
Migration is not always a smooth process. 
There are dangers of trafficking, sexual exploitation, and illegal transportation of migrants by smugglers across national borders \cite{gross2011migrations}. 

\subsection{Migration chain and regulation challenges}

Migration activity is not random and is believed to follow a systematic process via the so-called \textit{migration chain} involving a series of private processes and actors, e.g.~lawyers, agents, smugglers and migrants, who ensure smooth and continuous migration in return for monetary benefits \cite{castles2000international}. 
A basic form of pre-planned migration is initiated by a relative or friend residing at the intended destination who ensures the beginning-to-end success of the activity \cite{castles2013age}. 
Diverse actors in the chain are considered to obstruct efforts to regulate illegal migration. 
This challenge is further compounded by the widespread use of social media platforms and technological tools and services. 
For instance, \textit{smartphones}, which support all the required technological services\footnote{Services such as the Internet, Google Maps, Global Positioning System (GPS), Social Media Services and telephony}, proved to be an essential infrastructure in this regard \cite{gillespie2016mapping}. 
To prevent migration dangers and regulate the process, significant resources\footnote{For a detailed description of the European Agenda on Migration, see \url{ec.europa.eu/home-affairs/what-we-do/policies/europeanagenda-migration_en}} have been mobilised by the \textit{EU Member states} to tackle the problem. 

\subsection{Regulatory measures} 

A basic form of regulation is the categorisation of migrants as temporary labour migrants, highly-skilled and business migrants, irregular migrants/undocumented migrants, refugees, asylum seekers, and return migrants \cite{castles2000international}. Each group of migrants have different requirements, which can be used to regulate or prioritise their needs. For instance, a strict regulation may be placed on \textit{irregular migrants}, while \textit{asylum seekers} may have a higher priority.
Regulatory measures of migration are becoming more challenging to enforce. 

Firstly, the description of what constitutes \textit{migration governance} is considered contentious, due to the diverse definitions by different stakeholders. 
An inclusive definition sees migration governance as a type of global governance \cite{betts2010survival}:
\begin{quote}
    \textit{global governance includes a range of norms, rules, principles, decision-making procedures that exist over and above the level of a single nation-state}
\end{quote} 

\noindent The soundness of a migration governance can be assessed against the \textit{Migration Governance Indicators}\footnote{See: \url{gmdac.iom.int/migration-governance-indicators}}. 

\begin{figure}[!t]
\centering\includegraphics[width=.9\linewidth]{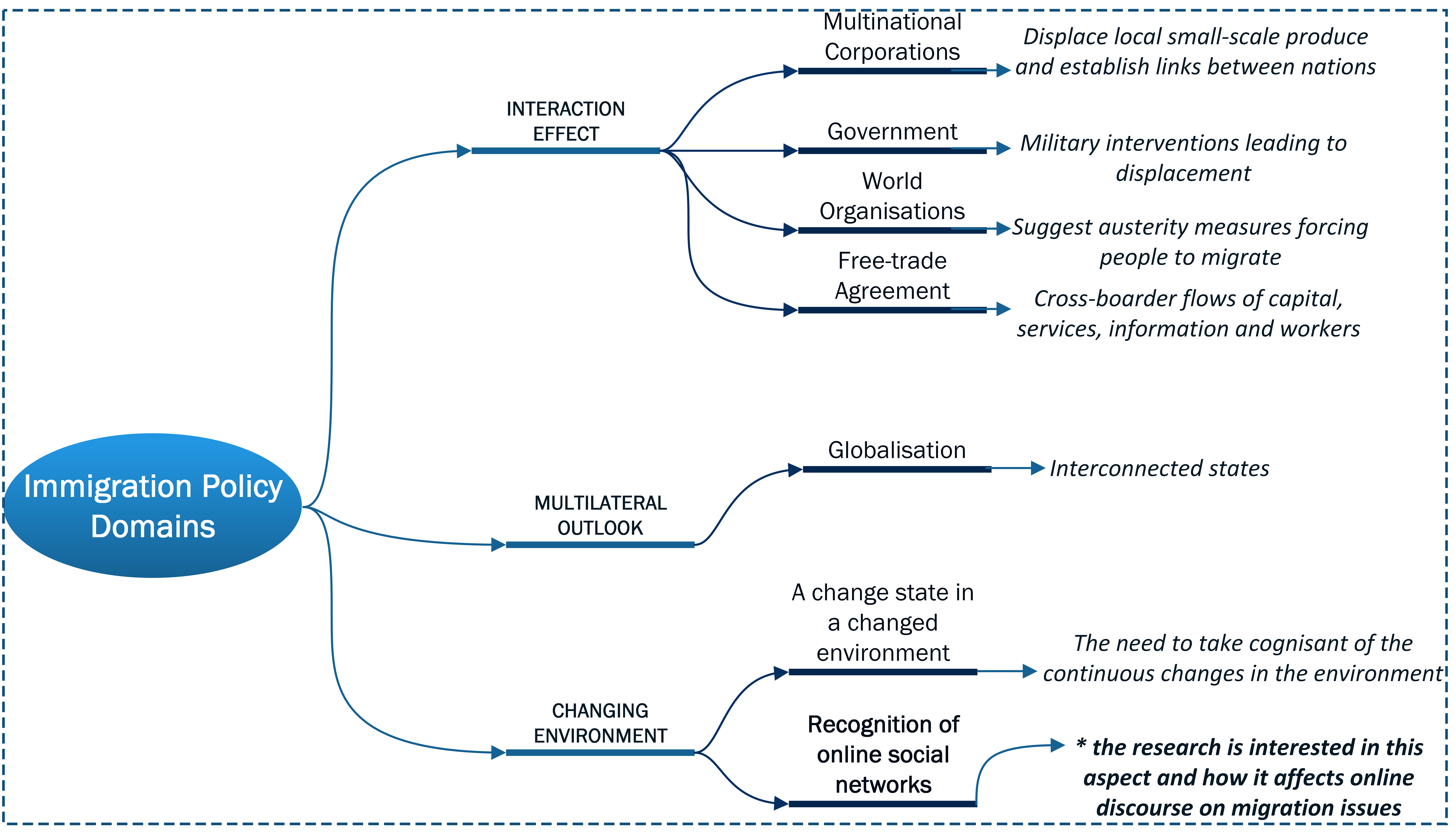}
\caption{Key elements in the formation of an effective immigration policy framework. 
         Each domain can be a potential trigger of migration.}
\label{fig1:migration-policy}
\end{figure}

The second challenge is \textit{globalisation}. 
The sovereignty of nations makes it possible for states to formulate and implement policies governing various affairs of the state and migration policies are no exception. 
However, globalisation is considered to place some restrictions and emphasises the consideration of many internal and external factors. 
\citeauthor{sassen2005regulating} \cite{sassen2005regulating} identifies major cardinals upon which conventional policies on migration are enacted from a unilateral perspective which recognises immigration policies as: 
(1) autonomous from other policy domains, 
(2) a solely sovereign matter, and 
(3) unaffected by domestic and international transformations. 
These principles are blurred due to interconnected economic systems, transnational processes and consideration of other stakeholders' interests. 
Globalisation warrants immigration policies with a multilateral outlook. 
The shift from \textit{unilateral} to \textit{multilateral} gains greater momentum through the adoption of the \textit{United Nations General Assembly (2015) Sustainable Development Goals} and corresponding targets concerning sustainable development on well-managed migration that recognises mainstream global development policy \cite{stuart2017transforming}. 
Figure \ref{fig1:migration-policy} summarises areas that need to be encompassed within migration policies \cite{sassen2005regulating}.

The \textit{changing environment domain} in Figure \ref{fig1:migration-policy} is concerned with how society is changing, especially with respect to social interactions. 
Recognising the relevance of the \textit{changing environment domain} in the Figure, we opined that insights from social media would support and inform a more inclusive strategy to address some of the regulatory challenges. 
We aim to complement existing regulatory measures within the broad area of integration by evaluating perceptions/opinions of the public using social media data.

\section{Related Work}
\label{sec3:review}

We review related work spanning the disciplines of computer science and social science. 
The focus is on prior research on \textit{sentiment analysis} from a computational perspective and from empirical social science studies to understand \textit{public opinions on migration}. 
To conclude, we discuss related work on \textit{migration and social networks}.

\subsection{Public opinion on migrants}

A 2016 report suggests that approximately 65.3 million people are displaced due to forced migration \cite{unhcr2016global}. 
Migration culminates in the creation of a complex and super diverse metropolis, which is regarded as a major obstacle to the formation of effective policies that meet the demands of such various groups \cite{vertovec2007super}. 
Satisfying the needs of a diversified community is essential for the development of public policies \cite{bilecen2018missing}, and public
opinion plays a vital role in shaping an immigration policy \cite{lahav2004public}. 
The opinion of the public can be regarded as a consequence of circumstances, i.e.~a change in response to the gravity of the situation. 
For instance, the occurrence of a spectacular event may lead to the expression of strong opinions. 
Concerning immigration, public opinion plays a vital role and requires critical analysis of diverse public data. 
There is a seam of literature offering an understanding of public opinion regarding minorities.
Coenders et al.~\cite{coenders2003majorities} and Semyonov et al.~\cite{semyonov2006rise} utilised the \textit{Eurobarometer data} to understand the attitude of the public toward migrants. 
The data is in the form of a survey, collected over a long period within EU Member states. 
The studies found that the majority of the public resents migrants, and such insight is being recognised in policy formulation. 
In relation to public attitude toward social policies, it is of significance to understand the link between the expressed sentiment and the context of expression \cite{clemencecarl}. 
In this study, we analyse both the opinion and the context of expression by examining various discussion topics from online data. 
Incorporating knowledge from multifaceted public opinions from a social media perspective to complement other empirical studies \cite{callens2015integration} will be useful.

\subsection{Social networks and migration}

Social networks are useful in establishing effective migration chains. 
Recently, there is a renewed interest in examining the impact and relationship between \textit{social networks} and \textit{migration}. 
Bilecen et al.~\cite{bilecen2018missing} provide a concise review of relevant studies in this direction. 
Chelpi-den Hamer and Mazzucato \cite{chelpi2010role} examines how migrants integrate in the receiving community and the role of social networks in providing support.  
Similarly, Komito \cite{komito2011social} and Dekker and Engbersen \cite{dekker2014social} examine the role of social media in migrants' social interaction, especially in reconnecting migrants to distant kinships. 
Our focus in this study is the analysis of opinion on, and the discussions about migrants and refugees on social media.

\subsection{Sentiment Analysis}

There has been growing interest in sentiment analysis across many domains due to the utility of social media as a resourceful repository of public opinion.
Social media platforms offer a rich source of longitudinal data for mining opinion and sentiment expressed by users \cite{pak2010twitter}. 
Using social media data, many social phenomena such as rumour, emotion, sentiment and propaganda have been successfully studied. 
Qazvinian et al.~\cite{qazvinian2011rumor} explored a large collection of Twitter data to detect tweets conveying rumour by utilising features related to \textit{lexical patterns}, \textit{parts-of-speech}, \textit{retweets}, \textit{post limits}, \textit{URLs} and \textit{hashtags}. 
Sentiment analysis deals with understanding the polarity of opinion, i.e.~positive, neutral or negative, and sometimes the context of expression \cite{chakraborty2016fashioning}. 
Ljube{\v{s}}i{\'c} and Fi{\v{s}}er \cite{ljubevsic2016global} utilised tweets annotated with location information and \textit{emoticons} to study the mood of users at different geographical locations. 
Bollen et al.~\cite{bollen2011modeling} examined the reaction of online users regarding socio-economic phenomena, such as election outcomes and how to quantify them using an extended six-dimensional psychometric instrument. 
W{\"u}rschinger et al.~\cite{wurschinger2016using} analyses the evolution of derogatory terms, e.g.~\textit{rapefugee}, \textit{rapeugee} and \textit{rapugee} as a fusion of \textit{rape} and \textit{refugee}, orchestrated by opponents of migrants and refugees on Twitter.
Recognising the contribution of social media in better understanding of our modern society \cite{clemencecarl} and the influence of public opinion in shaping immigration policies \cite{lahav2004public}, we present a comprehensive analysis of online public opinion about \textit{migrants} and \textit{refugees}. 

\section{Materials, Methods and Results}
\label{sec4:experiments}

\begin{figure}[t]
    \centering
    \includegraphics[width=.7\linewidth]{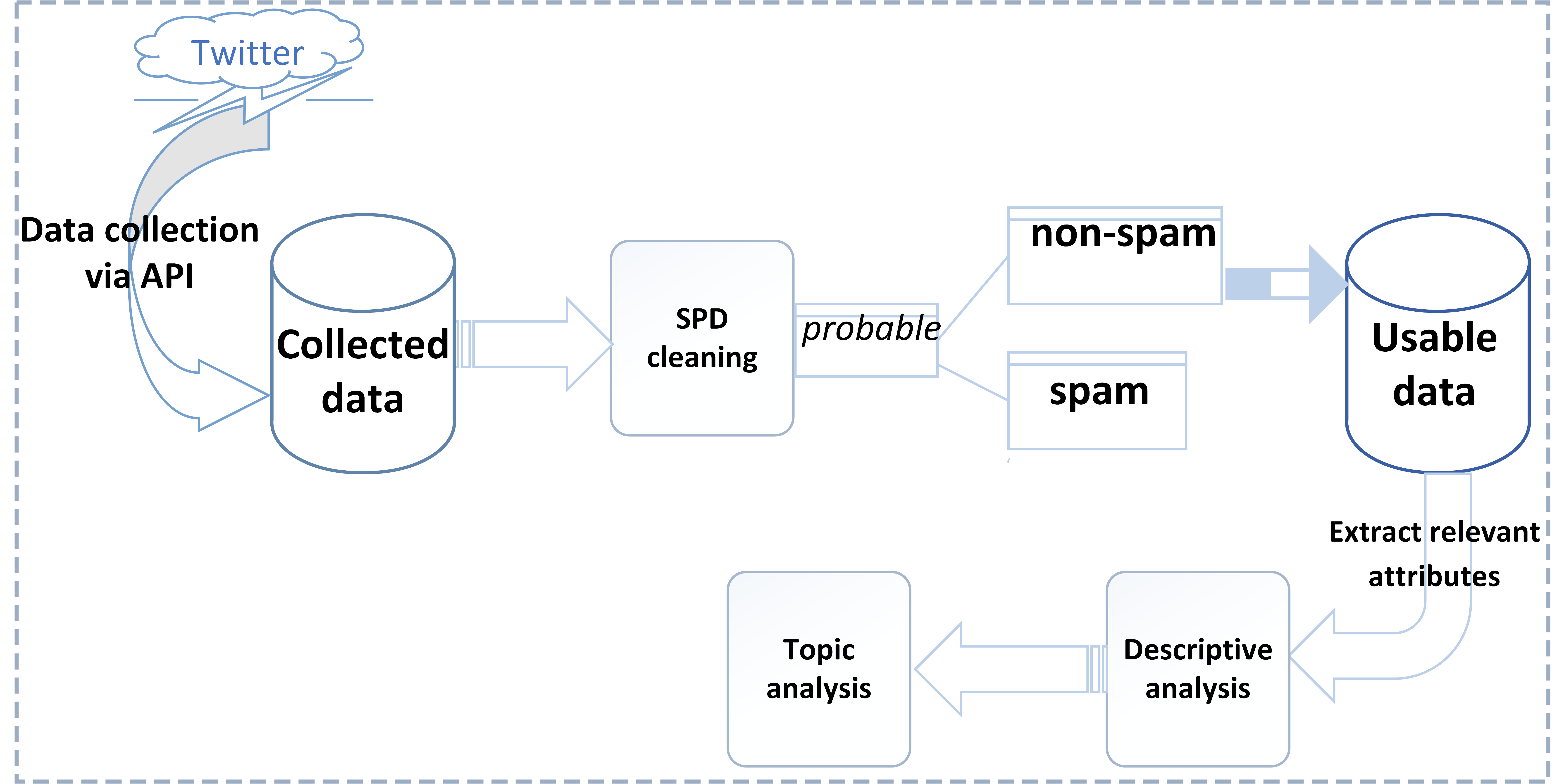}
    \caption{A high-level illustration of the research workflow, from data collection to analysis}
    \label{fig:process-workflow}
\end{figure}

In this section, we present the data collection method and describe our sentiment analysis method in detail. Figure~\ref{fig:process-workflow} shows the underlying architecture summarising our approach from raw data collection to the analysis of results. 

\subsection{Dataset: collection and preprocessing}

Twitter is an ideal avenue to collect data on different topics from a variety of individuals/users and locations. 
We chose to use Twitter due to the following facts: 
\begin{enumerate}[leftmargin=*,labelsep=4.9mm]
    \item Twitter is used by many diverse people to share various aspects of their social lives: fads, opinion, breaking news, etc. 
    \item There is huge amount of data readily available for analysing various social phenomena, such as migration in this study. 
    \item During the \textit{EU refugee crisis}, Twitter was actively utilised to report and discuss many related incidences. 
\end{enumerate}
The dataset for this study consists of a collection of short messages (known as tweets) posted by Twitter users during the height of the \textit{EU refugee crisis}, \textit{(2016/2017)}. 
Noting the bias that may arise due to the seemingly black box sampling strategy of Twitter to return queried documents \cite{tromble2017we}, we utilise different keywords covering various aspects of the subject. 
Keywords play a crucial role in retrieving specific documents from large corpora \cite{lott2012survey}.  
We collected tweets satisfying predefined searching criteria, shown in Table \ref{tab1:dataset}. 
About a third of the collected tweets have been discarded for being flagged as spam or irrelevant to the subject matter. 
Table \ref{tab1:dataset} summarises the dataset and shows some collection keywords.

\begin{table}[t]
  \caption{Dataset and collection keywords}
  \label{tab1:dataset}
  \centering
  \begin{tabular}{lp{8cm}}
  \toprule
    \textbf{Dataset}    & Refugee Data \\
    \textbf{Size}   & 0.8 milion tweets \\
    \multirow{2}{*}{\textbf{Description}}    & It consists of tweets collected in 2016/17 related to the EU refugee crisis. The original set consist of 1.4 million tweets. \\
   \textbf{Sample collection keywords}
        & \textit{refugee(s)}, \textit{migrant(s)}, \textit{``refugee crisis''}, \textit{crisis}, \textit{EU refugees}  \\
    \bottomrule
\end{tabular}
\end{table}

\subsubsection{Preprocessing} 

The ease of obtaining diverse data in a vast quantity also comes with challenges. 
Tweets are generally noisy and contain a substantial proportion of irrelevant or spam content. 
We utilise a spam filtering technique proposed by Inuwa-Dutse et al.~\cite{inuwa2018detection} to get rid of irrelevant content from the data as a first \textit{preprocessing} task. 
This was followed by data normalisation involving the removal of stopwords and the conversion of all text to lower case.

\subsection{Sentiment and topic analysis}

For the sentiment analysis task, we utilised a popular sentiment scoring tool, \textit{Valence Aware Dictionary and sEntiment Reasoner (VADER}\footnote{VADER is available as a sentiment scoring package in Python \cite{gilbert2014vader}.}), which has been trained on extensive collections of diverse datasets suitable for this task. 
We opted for \textit{VADER} instead of developing a custom sentiment analyser due to the following reasons:
\begin{itemize}[leftmargin=*,labelsep=5.8mm]
    \item VADER is well suited to work with a variety of data types including social media. 
    \item It is able to capture the nuances arising from a diversity of posting styles and is capable of detecting sentiment of different gravity and intensity. 
    \item It is trained on a huge amount of diverse datasets (more diverse and larger than our data) to understand the context and sentiment orientations of various lexicons. 
    \item It is recognised in prior research as an effective toolkit \cite{benamara2007sentiment,subrahmanian2008ava}.
\end{itemize}

\subsubsection{Sentiment in tweets} 

We compute the sentiment score for each tweet in the dataset. 
Each score is based on the proportions of \textit{negative}, \textit{neutral} and \textit{positive} sentiment which sum up to 1. 
For instance, a given tweet can be scored as $71\%$ \textit{positive}, $29\%$ \textit{negative} and $0\%$ \textit{neutral}. 
We also compute the compound score, \textit{s}, that measures the sentiment intensity and ranges from extremely negative (-1) to extremely positive (+1) for each tweet.
For instance, a tweet is considered positive if the compound score is greater than a threshold, e.g.~\textit{s} $\ge 0.05$ for a \textit{positive} or \textit{s} $ > -0.05$ to $0.05$ for a \textit{neutral} and \textit{s} $\le -0.05$ for a \textit{negative} tweet. 
\Cref{fig2:proportions of sentiment,fig3:sentiments: polarity and intensity,fig4a:sentiment: by region,fig4b:sentiment:regionalProportions,fig5a:sentiment: verified users,fig5b:sentiment: unverified users} show the sentiment analysis results on our dataset.

\subsubsection{Topic analysis}

In addition to sentiment analysis, we analyse the topics in each sentiment category using a topic modelling technique based on \textit{Latent Semantic Analysis (LSA)}. 
LSA is a useful decomposition technique that can be used to capture the semantics and relevance of a set of terms to a document \cite{wiemer2004latent}. 
Based on the semantics of the terms from LSA analysis, we group terms into themes and manually assign a high-level description of the meaning of each theme.  
Tables \ref{tab2:discussion themes} and \ref{tab3:discussion themes} in the appendix show some results. 
This is important in reinforcing and broadening our understanding of the opinions and the context of expressions.

\section{Discussion}
\label{sec5:discussion}

In this section, we present a detailed analysis of our results. 
We first explain the overall proportions of sentiment polarities in public opinion. 
This is followed by analysing sentiment based on  the \textit{geographical location} of users and their \textit{category} on Twitter, \textit{verified} or \textit{unverified}. 
We also describe the themes from \textit{LSA topic modelling} and how \textit{influential users} on Twitter can help in facilitating civilised and objective discussions on sensitive issues.

\subsection{Polarity in public opinion}

\begin{figure}[!t]
    \centering
    \includegraphics[width=0.35\linewidth]{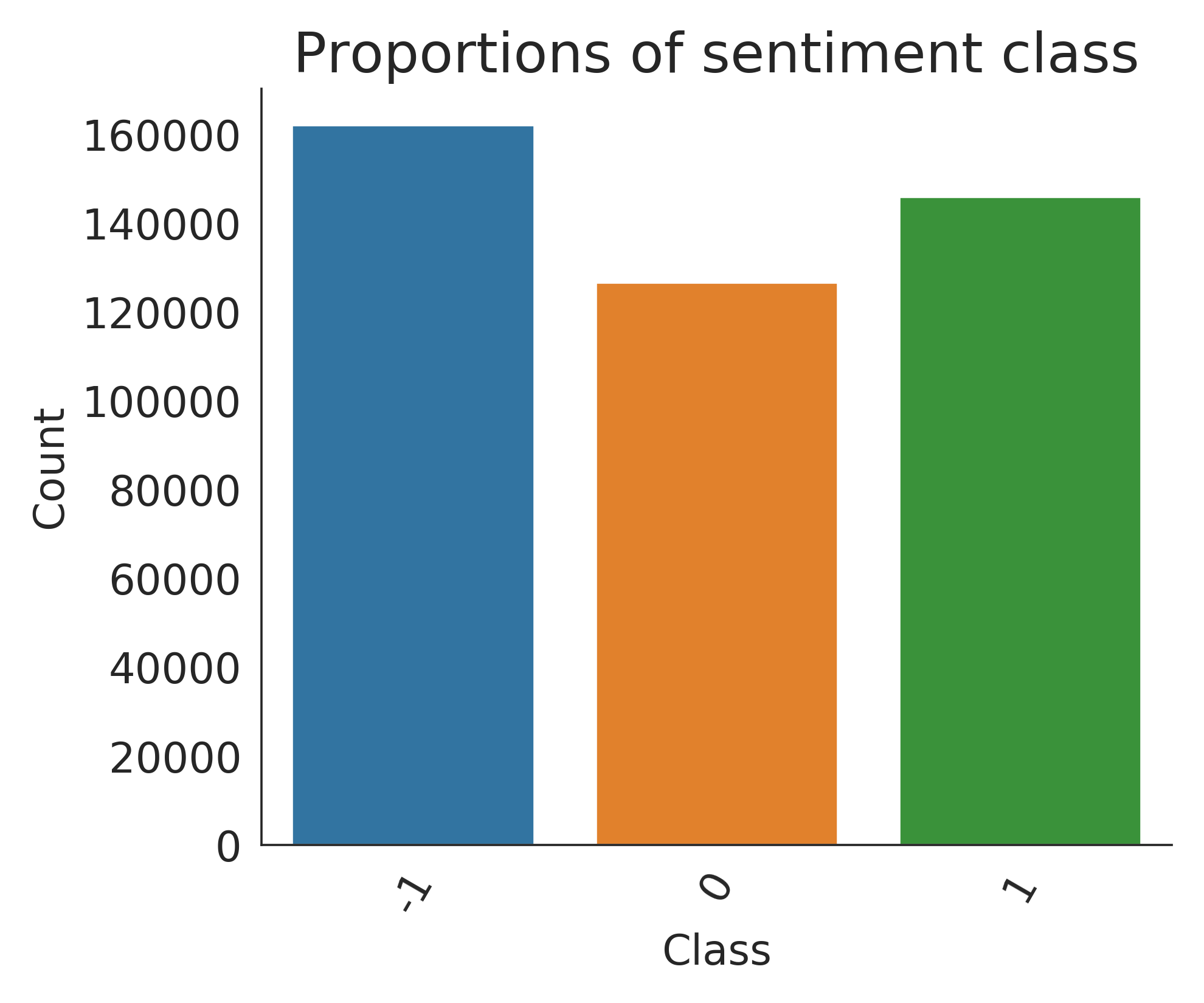}
    \caption{Number of tweets in the data set that correspond to the three sentiment polarities.}
    \label{fig2:proportions of sentiment}
\end{figure}

\begin{figure}[!t]
    \centering
    \includegraphics[width=.7\linewidth]{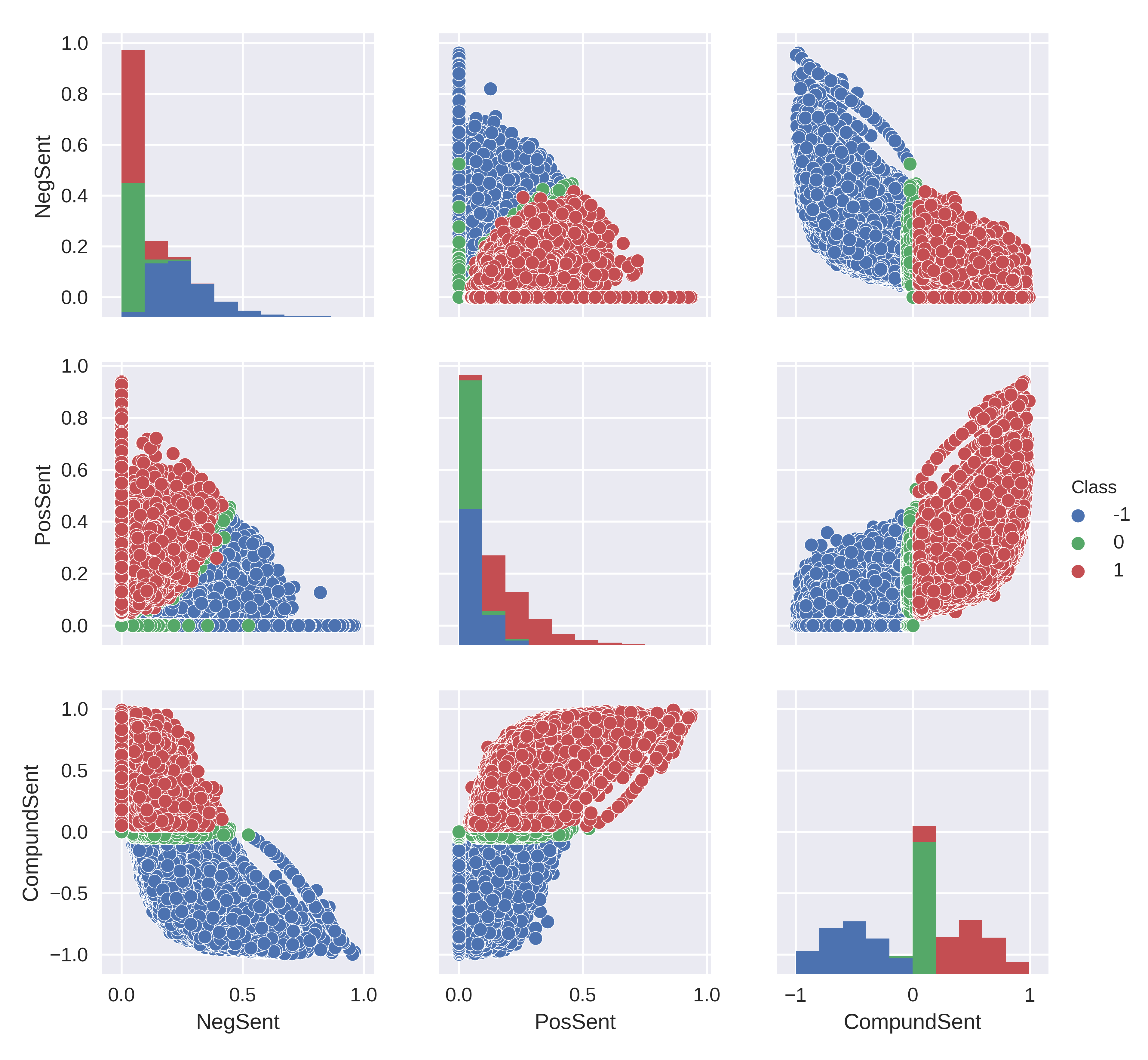}
    \caption{Proportions of sentiment polarities in tweets. 
             The proportions of \textit{negative} and \textit{positive} tweets are almost equal.}
    \label{fig3:sentiments: polarity and intensity}
\end{figure}

Figure \ref{fig2:proportions of sentiment} shows the proportions of sentiment polarities in all analysed public opinion. 
The difference between positive and negative sentiment is marginal. 
There is also a relatively high number of neutral tweets. 
The high proportion of positive sentiment may be due to \textit{influential} users of Twitter, such as celebrities and politicians, and credible news sources reporting incidences about migrants and refugees. 
Figure \ref{fig3:sentiments: polarity and intensity} shows a more elaborate distribution of sentiments in the public opinion.

\subsection{Sentiment by region and user group}

\begin{figure}[!t]
    \centering
    \includegraphics[width=.7\linewidth]{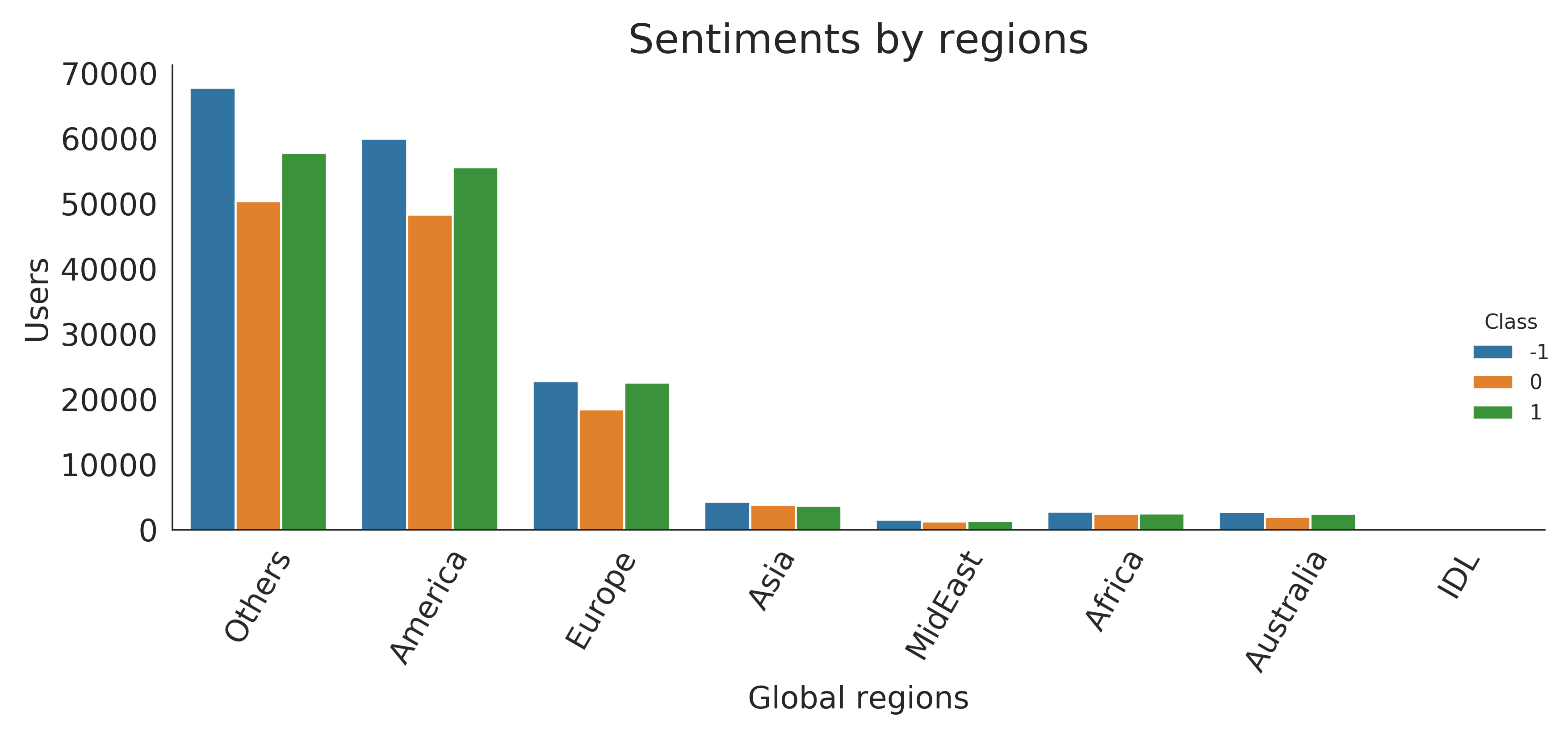}
    \caption{Sentiment polarities spanning across continents and group of users, i.e.~\textit{verified} or \textit{unverified}}
    \label{fig4a:sentiment: by region}
\end{figure}

\begin{figure}[!t]
    \centering
    \includegraphics[width=.7\linewidth]{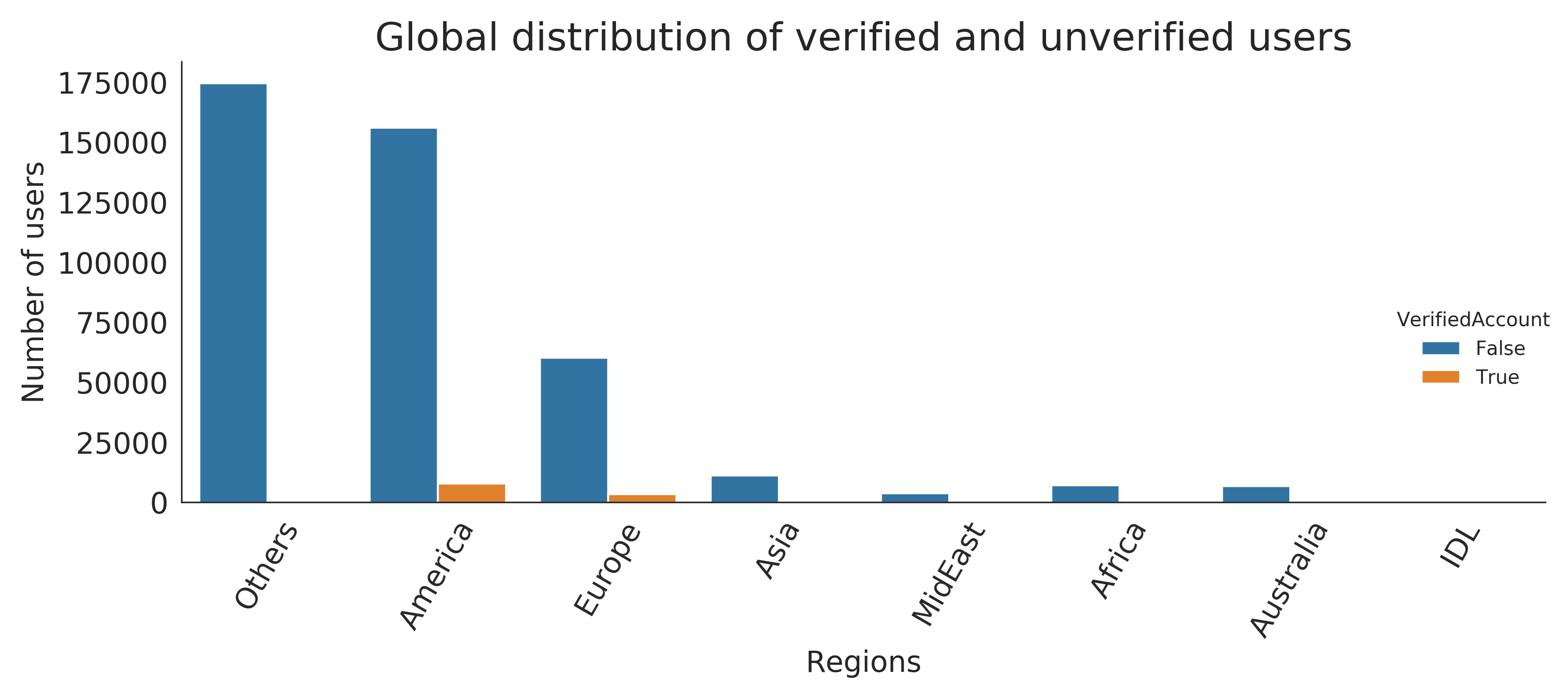}
    \caption{Sentiments from verified and unverified users across continental regions}
    \label{fig4b:sentiment:regionalProportions}
\end{figure}

Next we analyse the proportions of sentiment across the continents, as shown in Figure \ref{fig4a:sentiment: by region}, and based on the users' group, i.e.~\textit{verified} or \textit{unverified users}, as shown in Figure \ref{fig4b:sentiment:regionalProportions}. 
Verified users, such as politicians, celebrities and media personalities, tend to be socially influential and attract many followers on Twitter. 
Figure \ref{fig4a:sentiment: by region} shows that the main sources of opinion are from Northern America, Europe and Asia. 
Sentiment polarities in the opinion tweets from Asia, the Middle East, Africa and Australia are balanced in number. 
The proportion is slightly higher in favour of \textit{negative sentiments} in opinion tweets from North America and Europe.
The small numbers of opinion tweets from the Middle East and Africa partially corroborate an earlier report by \cite{smapp2016}, which examined the role of media in communicating the Syrian crisis to the rest of the world. 
The study observes that the majority of discussion on Syrian refugees takes place mostly in Syria and neighbouring countries, such as Jordan and Lebanon followed by Egypt and the Gulf states. 
Users in Maghreb countries, such as Morocco, Tunisia and Algeria, have posted a small number of tweets about the crisis.  This fact can be attributed to various local reasons, such as the availability and usage of internet services.

Previous studies based on the \textit{Eurobarometer data} found that aggression and negative views about migrants is more popular among the socially disadvantaged or low-income citizens due to competition for employment. 
This information cannot be confirmed solely based on online data. 
We follow a rather simplistic data-driven approach by assuming that most \textit{verified users} on Twitter have certain social advantages. 
A verified user account is certified by Twitter to be genuine, and the majority of account owners have a large number of followers, and thus exert a measure of influence. 
This is useful in understanding how such influential users encourage or discourage certain views or sentiment in relation to the subject matter. 
Figure \ref{fig4b:sentiment:regionalProportions} shows the proportions of sentiment polarities across different users groups, i.e.~\textit{verified} and \textit{unverified} users across continental regions. 

\begin{figure}[!t]
    \centering
    \includegraphics[width=.7\linewidth]{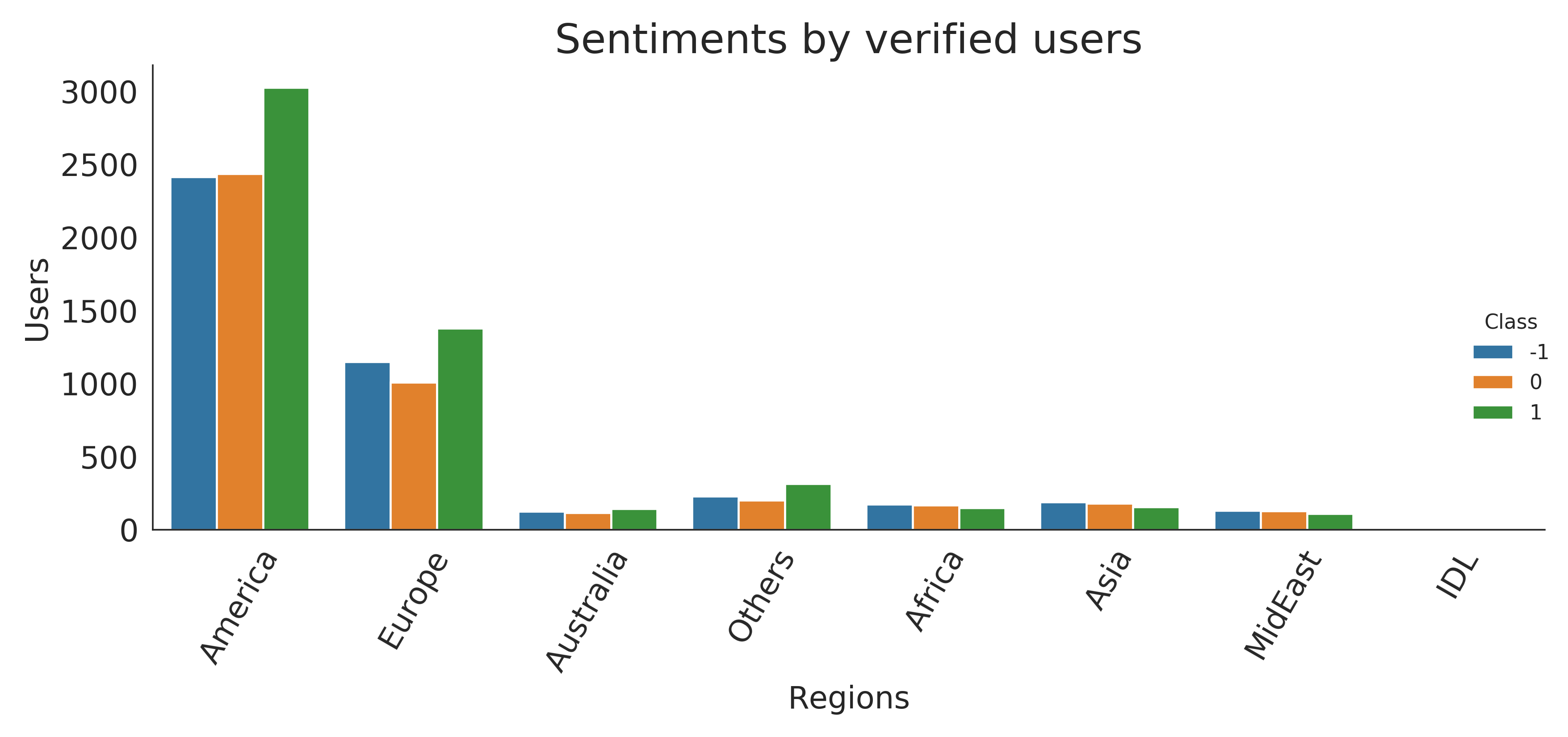}
    \caption{Verified users show low negative sentiments and generally higher positive sentiments.}
    \label{fig5a:sentiment: verified users}
\end{figure}

\begin{figure}[!t]
    \centering
    \includegraphics[width=.7\linewidth]{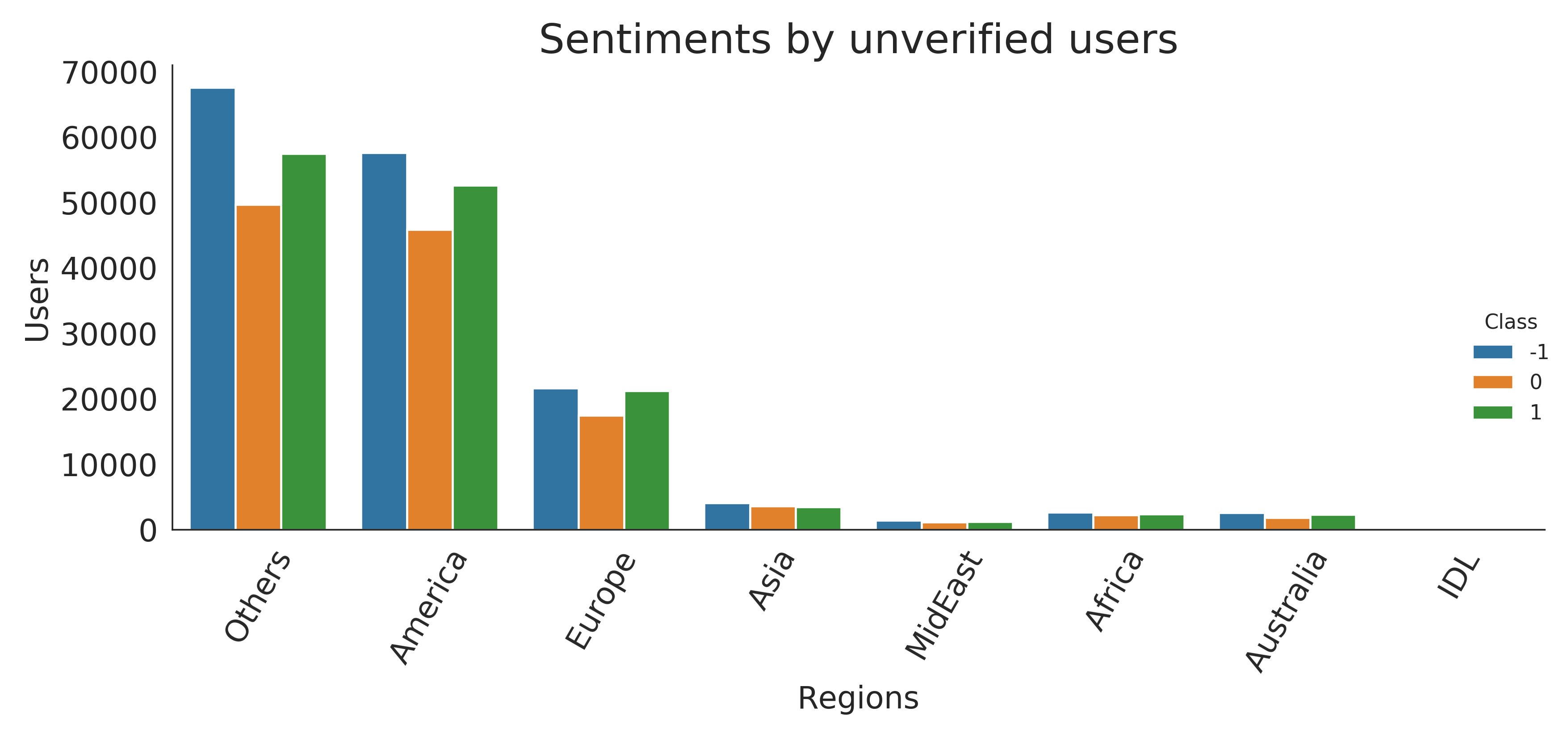}
    \caption{Substantial part of the negative sentiments comes from the unverified users group}
    \label{fig5b:sentiment: unverified users}
\end{figure}

Figure \ref{fig5a:sentiment: verified users} and Figure \ref{fig5b:sentiment: unverified users} show the sentiment analysis results on verified and unverified users, separately.
In the \textit{verified} user group, we observe a disproportionate proportion of \textit{negative} vs.~\textit{positive} with high proportion of neutral and positive sentiments on the issue. 
This is in contrast with \textit{unverified users} in which there is a significant number of negative sentiments in comparison to sentiments from \textit{verified users}.

The high number of positive sentiments in Figure \ref{fig5a:sentiment: verified users} can be attributed to how \textit{influential users} engage in the discussion. 
Tweets from such users are more likely to be captured during data collection, and such users tend to be sensitive or wary of expressing derogatory opinions. 
Secondly, credible news sources report incidences about migrants and refugees, which also tend to avoid being negative. 

\subsection{Role of influential users in neutralising negative opinions}

The term \textit{influential users} refers to users whose accounts have been \textit{verified} by Twitter. 
These users often have a high number of followers, which boosts their social status. 
Despite the small number of \textit{verified users} in the data (see Figure \ref{fig4b:sentiment:regionalProportions}), the proportion of positive sentiments is encouraging. 
This can be seen as a form of \textit{domino effect} where the \textit{verified users'} opinions resonate proportionally with their followers. 
We opine that verified users or any user with influence on social media can play a significant role in promoting civilised engagement on sensitive subjects affecting humans lives. 
The continuous engagement of influential users in this kind of discussion can help in neutralising negative online sentiment, which may result in a physical assault on minority groups in the receiving communities.
In contrast, some users express negative sentiments regarding migrants and refugees. 
These positive and negative sentiments are expressed at an equal rate. 
Neutralising the negative view can be useful in ensuring the peaceful integration of refugees and \textit{influential users} can play a pivotal role in this respect.

\subsection{Latent themes/topics}

Tables \ref{tab2:discussion themes} and \ref{tab3:discussion themes} in the appendix show some latent themes/topics, induced by \textit{LSA}.
According to the semantics of the terms, we group the collection of terms into themes under the following categories:
\begin{itemize}[leftmargin=*,labelsep=5.8mm]
    \item Table \ref{tab2:discussion themes} shows some prevalent latent themes from \textit{generic and neutral} sentiment polarities.
    This category consists of a combination of \textit{positive, negative and neutral} tweets. 
    Neutral tweets are indifferent and most of the time unrelated to the subject matter.
    \item  Table \ref{tab3:discussion themes} shows some latent themes from \textit{positive and negative} sentiments. 
    Themes in \textit{positive sentiment} are not as interpretable as their counterparts in \textit{negative sentiment} which are more understandable and derogatory. 
    Some meaningful topics in the \textit{positive sentiment} can be attributed to news reporting or appeals to support migrants and refugee groups.
\end{itemize}

\subsection{Key insights}

Insights from this study can be utilised to complement existing efforts toward a more inclusive migration regulatory framework. 
We focus on challenges associated with the smooth integration of migrants and refugees into society. 
 
Social media provide a valuable source of public opinion. 
Regular analysis from opinion mining of social media can better inform governmental decision making.
A recent joint effort by the European Economic and Social Committee and the European Commission in strengthening the integration of migrants recognises public perceptions as instrumental in ensuring smooth and peaceful coexistence between migrant groups and local communities \cite{eesc2018}. 
Incorporating knowledge from multifaceted public opinion from the social media perspective can complement previous studies \cite{callens2015integration,semyonov2006rise}. 
For instance, the analysis output from Twitter will be of relevance in part to the European Monitoring Centre on Racism and Xenophobia (\textit{EurWORK})\footnote{See: \url{eurofound.europa.eu/observatories/eurwork/industrial-relations-dictionary/european-monitoring-centre-on-racism-and-xenophobia}}. 
With readily available social media data, perceptions of the public can be captured as demonstrated in this study. 
These will ultimately help in delivering a comprehensive/holistic regulatory framework that recognises both offline and online public opinion. 
It requires a sound understanding of the diverse perspectives of migration by the public, the Government and other stakeholders to deliver effective regulation of the safety, rights and peaceful coexistence of immigrants with the receiving communities.
    
\section{Conclusions}
\label{sec6:conclusion}

Arguably, containing or regulating the rate of migration and controlling the level of resentment towards migrants is one of the pressing issues of modern civilisation.
Discourse on migrants is a major public issue causing much controversy among politicians and the general public due to the political consequences of the issue and various reports from the media. 
Decades of offline data collected via traditional survey methods have been utilised to understand public opinion and sentiment on immigrants. 
Earlier studies based on the \textit{Eurobarometer} data have shown a high proportion of opposing or conflicting views from the public. 
Settling and integration in the host communities may be challenging due to cultural diversity and ethno-cultural differences leading to various opinion views. 
Public opinion has been shown to influence and shape immigration policies. 
Diversifying the source of information to include social media platforms, such as Twitter and Facebook, is crucial toward informing/proffering effective regulation and reflecting the \textit{changing environment}. 

In this study, we analysed online public opinion and discussion topics on migration and refugees. 
We also analysed sentiments according to the geographical location of the posting user and according to user categories. 
Negative sentiment is dominant on the subject matter. 
There is a higher number of relevant tweets from North America and Europe, whereas numbers are lower for Asia, Africa, the Middle East and Australia. 
We examined how different users express opinion and concluded that \textit{verified} users express a less negative sentiment than \textit{unverified} users. 
Our study demonstrates the usability of social media as a complementary source of data in tackling challenges related to migration and refugees. 
This ensures a sound understanding of the problem by recognising diverse perspectives and concerting efforts for a deeper understanding of the issue. 
The dataset utilised in this study is small compared with the entire Twitter ecosystem. 
However, the sample size was designed to give depth rather than breadth and is of sufficient size to allow extracting meaningful conclusions \cite{tight2006research}.
By continuously harvesting and analysing the data, a higher and more in-depth understanding of the challenges of large scale migration will result in improved strategies to address the issues.

\vspace{6pt} 

\authorcontributions{
conceptualization, I.I.D. and I.K.; 
methodology, I.I.D. and I.K.; 
software, I.I.D.; 
validation, I.I.D.; 
formal analysis, I.I.D.; 
investigation, I.I.D. and I.K.; 
resources, I.I.D.; 
data curation, I.I.D.; 
writing--original draft preparation, X.X.; 
writing--review and editing, I.K. and M.L.; 
visualization, I.I.D. and I.K.; 
supervision, M.L. and I.K.; 
project administration, I.K.; 
funding acquisition, I.K.
}

\funding{The third author has participated in this research work as part of the TYPHON Project, which has received funding from the European Union's Horizon 2020 Research and Innovation Programme under grant agreement No.~780251.}

\acknowledgments{The authors would like to thank Prof.~Francesco Rizzuto for the fruitful discussions and exchange of ideas about a multitude of aspects related to social media, spam content and the motives of spammers.}

\conflictsofinterest{The authors declare no conflict of interest.} 

\appendixtitles{yes}
\appendix
\section{Analysing discussion topics}

Table \ref{tab2:discussion themes} and Table \ref{tab3:discussion themes} show some examples from the \textit{topic analysis} results. 
Terms appearing in the majority of tweets in the collection are grouped together as latent themes/topics. 
The \textit{Remark} column in the tables is based on our manually-assigned estimate of the likely context and interpretation of the corresponding theme/topic. 
Our remark for each theme is in the form of \textit{news reported}, \textit{appeal} or \textit{campaign} on behalf of migrants/refugees, \textit{trivial}, \textit{pejorative} or \textit{unrelated}. 

\begin{table}[H]
\begin{center}
  \caption{Examples of themes in generic and neutral tweets}
  \centering
  \label{tab2:discussion themes}
  \begin{tabular}{ccll}
   \toprule
    & \textbf{Id}    & \textbf{Themes} & \textbf{Remark}\\ 
    \midrule
    \multirow{17}{*}[-0.4ex]{\begin{sideways}\textbf{Generic}\end{sideways}}
        & t0  & [refugees, migrants, via,syrian, trump, take]  & news   \\
        & t1  & [people, take, help,immigrants, realdonaldtrump,crisis]  & appeal   \\
        & t2  & [refugee, trump, ban, children, court, camp]  & news   \\
        & t3  & [ban, court, travel, supreme, muslim, administration]  & news   \\
        & t4  & [syrian, refugees, libya, markets, slave, vulnerable]  & news   \\
        & t5  & [people, libya, african, country, refugee, markets]  &    \\
        & t6  & [new, welcome, york, crisis, need, migrants]  & appeal   \\
        & t7  & [news, country, germany, fake, facebook, child]  &   \\
        & t8  & [children, like,asylum, women, canada, refugees]  &   \\
        & t9  & [like, europe, welcome, news, sweden, realdonaldtrump]  &    \\
        & t10 & [realdonaldtrump, germany, new, merkel, safe,shelter]  &   \\
        & t11 & [eu, ban, country, us, court, travel]  & \\
        & t12 & [realdonaldtrump, news, safe, fake, refugeeswelcome, shelter]  &   \\
        & t13 & [jews, africans, nativity, scene, arabs, refugee]  & pejorative   \\
        & t14 & [starved, coast, attack, somali, migrants, murder]  &   \\
        & t15 & [war, syrian, home, forced, flee, persecution]&   \\
        & t16 & [obama, un, great, rohingya, video, war]    & news   \\
    \midrule 
    \multirow{7}{*}[-0.4ex]{\begin{sideways}\textbf{Neutral}\end{sideways}}
        & t0 & [cgv0cjfbvx, h8xbrdx21a, prexy, scarey, presidenttrum, tom\_lewisville]  & trivial   \\
        & t1 & [md0bdfzwdb, mickek69, energy, pnlli03hil, carl0sperdue, unionised]  & trivial   \\
        & t2 & [president, trum, potatoehead, stampa\_estera, caliph, h8xbrdx21a]  & pejorative   \\
        & t3 & [medicals, alt, ramadanbombathon, a5kkxvosgf, incorruptible, unclesam]  &    \\
        & t4 & [stampa\_estera, cks0wfxeuz, atipconsulting, salaries, adrianh33184399, ridiculously]  & unrelated   \\
        & t5 & [securitising, malkamelanie, caliph, scarey, qfzy62iiu7, o2qpqez3vv]  &   \\
        & t6 & [editors, africans, kcstar, ckhiqc4z5j, terryja43095721, powerfulteachers]  & unrelated   \\  \bottomrule 
    \end{tabular}
\end{center}
\end{table}

\begin{table}[H]
\begin{center}
  \caption{Examples of themes in positive and negative tweets}
  \label{tab3:discussion themes}
  \begin{tabular}{ccll}
   \toprule
        & \textbf{Id}    & \textbf{Themes} & \textbf{Remark} \\ 
    \midrule
    \multirow{9}{*}[-0.4ex]{\begin{sideways}\textbf{Positive}\end{sideways}}
        & t0  & [mennonites,unhcrsom, lobbyists, paralized, godrockin, boils] & trivial \\
        & t1  & [unhcrsom, suntimes, russianpawn, secretcinema, hateus, paralized] & \\
        & t2  & [applies, spaced, paralized, ruth, unhcrsom, sosforyusuf] & appeal \\
        & t3  & [lcwr, spaced, unite, akraghawan, ralphshields313, corrupti] & \\
        & t4  & [rather, mennonites, cbus, fuckfacism, stephenhawking,cone] & pejorative \\
        & t5  & [livetodine, unhcrsom, jimsavege, sosforyusuf, godrockin, bolivia] & appeal \\
        & t6  & [livetodine, suntimes, girlfriend, malnutrit, randomactsofkindnessday] & appeal \\
        & t7  & [stick, garlic, randomactsofkindnessday, wa5pi9iffg, learntotrade, clipping] & appeal \\
        & t8  & [iledchat, jimsavege, nasrani, sunnyradio, erect, nationalistic] & unrelated\\ 
    \midrule
    \multirow{22}{*}[-0.4ex]{\begin{sideways}\textbf{Negative}\end{sideways}}
        & t0  & [ban, trump, people, country, immigrants, travel] & news \\
        & t2  & [migrants, libya, markets, stop, slave, slaves] &  \\
        & t3  & [ban, trump, travel, muslim, justices, unlawful] & news \\
        & t4  & [stop, crisis, markets, slaves, close, libya] & \\
        & t5  & [news, fake, syrian, facebook, court, selfie] & \\
        & t6  & [via, c0nvey, muslim, news, germany, attack] & news \\
        & t7  & [news, migrants, fake, facebook, syrian, refugee] & \\
        & t9  & [attack, people, migrants, berlin, yemen, killed] & \\
        & t11 & [syrian, children, women, raped, libyan,beaten] &  \\
        & t12 & [germany, eu, merkel, europe, travel, ban] & \\
        & t13 & [sweden, rape, trump, people, right, women] & \\
        & t14 & [us, syrian, want, illegal, canada, immigrants] & \\
        & t15 & [illegal, trump, eu, stop, le, pen, home] & \\
        & t16 & [australia, stabbed, policy, europe, cruelty, refugees, sinister] & \\
        & t17 & [new, murdered, war, crisis, girl, terrorist] & \\
        & t18 & [fuck, realdonaldtrump, dead, white, merkel, mediterranean] & pejorative \\
        & t19 & [muslim, immigrants, canada, policy,children, blind] & \\
        & t20 & [climate, military, unimaginable, stir, germany, aliens] & pejorative \\
        & t21 & [france, school, afghan, arrest, country, muslim] & \\
    \bottomrule
    \end{tabular}
\end{center}
\end{table}

\reftitle{References}
\externalbibliography{yes}
\bibliography{online_perceptions}

\end{document}